\documentclass[pra,aps,twocolumn,10pt,showpacs,nofootinbib]{revtex4-1}
\usepackage[utf8]{inputenc}
\usepackage[T1]{fontenc}
\usepackage{amsmath}
\usepackage{amsfonts}
\usepackage{amssymb}
\usepackage{mathcmd}
\usepackage{graphicx} 
\usepackage{dsfont}
\usepackage{braket}
\usepackage{todonotes}
\usepackage{bm}
\usepackage{upgreek}
\usepackage{bbm}
\usepackage{mathrsfs}

\newcommand{\mi}{\mathrm{i}}

\begin{document}

\title{Holonomic Gates in Pseudo-Hermitian Quantum Systems}
\author{Julien Pinske}
\author{Lucas Teuber}
\author{Stefan Scheel}
\email{stefan.scheel@uni-rostock.de}
\affiliation{Institut f\"ur Physik, Universit\"at Rostock, 
Albert-Einstein--Stra{\ss}e 23-24, D-18059 Rostock, Germany}

\date{\today}

\begin{abstract}
The time-dependent pseudo-Hermitian formulation of quantum mechanics allows
to study open system dynamics in analogy to Hermitian quantum systems. In this setting, we show that
the notion of holonomic quantum computation can equally be formulated for
pseudo-Hermitian systems. Starting from a degenerate pseudo-Hermitian Hamiltonian we show
that, in the adiabatic limit, a non-Abelian geometric phase emerges which realizes a pseudounitary
quantum gate. We illustrate our findings by studying a pseudo-Hermitian gain/loss system which can be
written in the form of a tripod Hamiltonian by using the biorthogonal representation. It is shown that this system 
allows for arbitrary pseudo-$\mathrm{U}(2)$ transformations acting on the dark subspace of the system.
\end{abstract}
\maketitle

\section{Introduction}

In the standard formulation of quantum mechanics (QM), observables are 
associated with Hermitian operators.
This Hermiticity condition ensures that the spectrum of the observable is 
real-valued, thus making a physical interpretation possible.
It was first shown by Bender and Boettcher~\cite{Bender} that also 
non-Hermitian systems, obeying $\mathcal{PT}$-symmetry (parity-time-reversal 
symmetry), can also show real spectra.
This observation revived serious investigations into unconventional quantum mechanics.
In particular, pseudo-Hermitian QM~\cite{Moste2} (and the related biorthogonal QM~\cite{Brody}) have received special attention.
This theory investigates pseudo-Hermitian systems, in which the Hamiltonian of the quantum system is non-Hermitian but can still be associated with a Hermitian counterpart.
Such peculiar behaviour leads to a whole class of new Hamiltonians that could
reveal interesting new physics.

In this work, we are particularly interested in the paradigm of holonomic 
quantum computation (HQC)~\cite{HQC,Pachos2}, which is based on the emergence 
of a (non-Abelian) geometric phase (holonomy) during a cyclic time evolution of 
a quantum system~\cite{Berry,Wilzeck}. Corresponding to a holonomy there is a non-Abelian
gauge field mediating the computation in form of a parallel transport. These types 
of gauge fields are realized in systems where the demand for degeneracy can be
satisfied.
Examples of such systems are cold atomic samples~\cite{Dalibard} or 
artifical atoms in superconducting circuits~\cite{Berger}. Recently, the implementation
of such gauge fields were realized in systems of coupled waveguides~\cite{Teuber}.
Another succesful scheme utilized the spin-orbit coupling of polarized light in 
asymmetric microcavities~\cite{MaSO}.

Holonomic quantum computing is a purely geometric approach to quantum 
computational problems. Unitary gates are implemented by generating a 
suitable holonomy from a Hamiltonian system. The transformation that a quantum 
state undergoes is the shadow (horizontal lift) of a loop in a parameter space 
(manifold) $\mathscr{M}$. In this context, the question of computational 
universality can be understood as the capability of generating a set of closed 
paths such that the holonomy spans up the entire unitary group~\cite{HQC}. 
Universality is typically reached only in a subspace of the whole Hilbert space 
$\mathscr{H}$, the so-called quantum code $\mathscr{C}$. The most common 
choice is to take $\mathscr{C}$ as the ground state of the Hamiltonian. This 
results in a type of ground-state computation in the lowest energy 
eigenvalue manifold~\cite{Pachos2}. The elements of the (quantum) code $\mathscr{C}$ are 
called the (quantum) code words, as the gates act on them and, in that way, 
perform the computation.  

Holonomy groups often appear in the context of gauge theories. This stems from 
their intrinsic connection to gauge fields, which can be elucidated by studying 
the theory of fiber and vector bundles~\cite{Bohm,Moste}. Physical 
implementations of holonomic gates were considered in non-linear Kerr media~\cite{Pachos}, superconducting quantum dots~\cite{QEC} or quantum 
electrodynamical circuits~\cite{Romero}, but to the best of our knowledge only 
for Hermitian systems. This broad range of possible implementations, together 
with the fault tolerance of HQC~\cite{Oreshkov}, make it desirable to generalize the 
concept of holonomic gates beyond Hermitian QM. 

It has been pointed out that, in order to generate a non-Abelian geometric 
phase (holonomy), the Hermiticity of the Hamiltonian is not a necessary 
condition~\cite{Bhandari}. Indeed, an explicit calculation of an Abelian 
geometric phase for a $\mathcal{PT}$-symmetric system has been provided in 
Ref.~\cite{Wang0}. However, because degeneracy plays such a crucial role in the 
theory of HQC, we will extend the theory from Refs.~\cite{Wang0,Wang,Gong2019} to the 
non-Abelian case. With this, one is in principle able to implement quantum 
computational gates by means of pseudo-Hermitian systems. The conservation of 
the norm of quantum states is of utmost importance and will be discussed in 
this work, referring to time-dependent models for pseudo-Hermitian QM. The 
occurrence of new physical effects from these types of holonomic gates is 
deeply connected to the question of measurable consequences of the underlying 
Hilbert space metric~\cite{Brody2}. The idea of a pseudo-Hermitian 
representation of geometric phases could also be of interest in the theory of 
open quantum systems. The latter subject showed, by studying lossy systems, 
deep relations to pseudo-Hermitian and $\mathcal{PT}$-symmetric QM.

This article is organized as follows. In Sec.~\ref{sec:dynamics} we briefly 
review the dynamics of pseudo-Hermitian systems~\cite{Wang2,Moste}, and 
emphasize the change of the Hilbert space metric associated with a 
pseudo-Hermitian quantum system~\cite{Heiss}. In this framework, we will show 
in Sec.~\ref{sec:holonomy} that it is possible to derive a non-Abelian gauge 
field arising from an adiabatic mapping onto a degenerate subspace of the 
system, by extending the ideas of Ref.~\cite{Wang} to the degenerate case. 
Sec.~\ref{sec:theory} contains additional remarks and theoretical 
considerations on the construction of pseudo-Hermitian Hamiltonians from a 
gain/loss system or a biorthogonal basis. Following that, we will discuss the 
example of a degenerate interaction Hamiltonian, whose Hermitian analogue can 
be found in the area of light-matter coupling.
The gauge field is explicitly calculated and properties of the system are discussed in detail in Sec.~\ref{sec:dark}. Finally, we summarize our results with some concluding 
remarks in Sec.~\ref{sec:conclusions}. In Appendix~\ref{sec:gauge}, we derive 
the transformation law for the gauge field. A more sophisticated treatment of 
the geometry of pseudo-Hermitian quantum systems involves Grassmann and Stiefel 
manifolds, which can be found in Appendix~\ref{sec:Grassmann}.

\section{Dynamics of pseudo-Hermitian systems}
\label{sec:dynamics}

We begin by briefly recalling the time-dependent dynamics of pseudo-Hermitian 
quantum systems, following mainly Refs.~\cite{Wang2,Wang,Gong2019,Moste}. We consider a time-dependent $N$-dimensional ($N<\infty$) pseudo-Hermitian Hamiltonian 
$H(t)\neq H^\dagger(t)$, that is, $\mathscr{H}\cong\mathds{C}^N$. The 
generalization to infinite dimensional systems might be well possible,
but is of marginal interest for HQC. Such a pseudo-Hermitian 
system can be viewed as being Hermitian with respect to a similarity 
transformation
\begin{equation}
\label{eq:1}
H^\dagger(t)=\eta(t)H(t)\eta^{-1}(t),
\end{equation}  
where $\eta(t)$ (we sometimes suppress the time argument for 
brevity) is the so-called Hilbert-space metric \cite{Bender2,Heiss}. The latter 
induces a new inner product
\begin{equation}
\label{eq:2}
\braket{\varphi,\psi}_{\eta}=\bra{\varphi}\eta\ket{\psi},
\end{equation}
for all vectors $\varphi,\psi$ in the new Hilbert space 
$\mathscr{H}_{\eta(t)}$. Note that Hermitian  operators 
in $\mathscr{H}$ do not have to be Hermitian in  $\mathscr{H}_{\eta(t)}$. 

A different point of view can be taken by investigating the eigenvalue problem 
of $H$ \cite{Brody}. For a Hermitian operator over $\mathscr{H}_{\eta(t)}$, all 
its eigenvalues are real and its instantaneous eigenstates
\begin{gather}
H(t)\ket{\phi_{n}(t)}=E_{n}\ket{\phi_{n}(t)},\nonumber\\
H^\dagger(t)\ket{\tilde{\phi}_{n}(t)}=E_{n}\ket{\tilde{\phi}_{n}(t)},
\label{eq:3}
\end{gather} 
form a biorthogonal basis 
$\{\ket{\phi_{n}},\ket{\tilde{\phi}_{n}}\}$ with
$\braket{\tilde{\phi}_{n}|\phi_{m}}=\delta_{nm}$ 
\cite{Moste}. Combining Eqs.~(\ref{eq:1})--(\ref{eq:3}), we find that 
$\ket{\tilde{\phi}_{n}}=\eta\ket{\phi_{n}}$.

The time evolution $U:\mathscr{H}\rightarrow\mathscr{H}$ of a quantum 
system differs from conventional QM in that $U$ is no longer unitary, 
$U^\dagger U\neq\mathds{1}$. However, as it was shown in Ref.~\cite{Wang2}, a 
generalized unitarity condition can be established. For any two physical states 
$\ket{\Phi(t)}=U(t,t_{0})\ket{\Phi(t_{0})}$ and 
$\ket{\Psi(t)}=U(t,t_{0})\ket{\Psi(t_{0})}$ in $\mathscr{H}$ one demands that
\begin{equation}
\label{eq:4}
\frac{\mathrm{d}}{\mathrm{d}t}\braket{\tilde{\Phi}|\Psi}
=\frac{\mathrm{d}}{\mathrm{d}t}\bra{\Phi}\eta\ket{\Psi}=0.
\end{equation}
Equation~(\ref{eq:4}), together with Eq.~(\ref{eq:1}), implies a generalized 
time-dependent Schr{\"o}dinger-like equation \cite{Wang2,Moste}
\begin{equation}
\label{eq:5}
\mi\frac{\mathrm{d}}{\mathrm{d}t}\ket{\Psi(t)}
=\Lambda(t)\ket{\Psi(t)},
\end{equation} 
where $\Lambda(t)$ is the generator of time-displacement given by
\begin{equation*}
\Lambda(t)=H(t)+\mi K(t),
\end{equation*}
with $K(t)=-\eta^{-1}(t)\dot{\eta}(t)/2$. 
Replacing the state vectors in Eq.~(\ref{eq:4}) by their time evolution 
$U(t,t_{0})=\boldsymbol{\hat{\mathrm{T}}}\mathrm{exp}\left(-\mi\int_{t_{0}}^{t}\Lambda(\tau)\mathrm{d}\tau\right)$ ($\boldsymbol{\hat{\mathrm{T}}}$ denotes time ordering)
and using Eq.~(\ref{eq:5}) one obtains
\begin{equation}
\label{eq:7}
\mi\dot{\eta}=\Lambda^\dagger\eta-\eta\Lambda,
\end{equation}
where the dot denotes the time derivative.
Equation~(\ref{eq:5}) can be rewritten conveniently by introducing a covariant 
derivative $\mathrm{D}_{t}=\mathrm{d}/\mathrm{d}t-K(t)$. We thus find
\begin{equation}
\label{eq:8}
\mi\mathrm{D}_{t}\ket{\Psi(t)}=H(t)\ket{\Psi(t)}.
\end{equation}

We conclude this section by highlighting the physical consequences of the 
dynamical model presented here. Note that we imposed the Hermiticity condition 
(under the metric $\eta$) for all times $t$ [cf. Eq.~(\ref{eq:1})]. 
For this to be true, the Schr\"odinger equation of conventional QM has to be 
replaced by the Schr\"odinger-like equation (\ref{eq:5}) to satisfy the 
unitarity condition, Eq.~(\ref{eq:4})~\cite{Wang2}. If one wants to 
retain the original Schr\"odinger equation $\mi\ket{\dot{\Psi}}=H\ket{\Psi}$, 
then Eq.~(\ref{eq:1}) is violated whenever the metric becomes time-dependent. 
This can be seen by replacing $\Lambda$ by $H$ in Eq.~(\ref{eq:7}). In this 
case, $H$ would no longer be an observable for times $t>t_{0}$~\cite{Moste}. 
Up until now, this seems to be not fully understood, and a number of different 
approaches to handle this problem were made. However, the model presented here 
does not produce any contradiction with conventional quantum mechanics and, as 
it was shown in Ref.~\cite{Wang2}, a proper mapping to conventional QM is 
possible. We expect that, however the final formulation of pseudo-Hermitian QM 
might look like, it will embody the physical demands made in this model up to a 
matter of notation.

\section{Derivation of the holonomy}
\label{sec:holonomy}

The occurrence of non-Abelian geometric phases (holonomies) is, in terms of 
differential geometry, associated with a connection, i.e. a unique 
separation of the (tangent) Hilbert space 
$\mathscr{H}=\mathscr{H}_{\mathrm{exc}}\oplus\mathscr{H}_{0}$ into an
$n_{0}$-fold degenerate ground-state subspace $\mathscr{H}_{0}$ and the space 
$\mathscr{H}_{\mathrm{exc}}$ containing all excited states. Such a separation 
can be technically realized by a gauge field (a local connection one-form). 
Because a dynamical systems leads in general to a time-dependent Hilbert space, 
we demand that this separation holds while the quantum states undergo a time 
evolution during the period $T$. Thus, any initial preparation 
$\ket{\Psi(0)}\in\mathscr{H}_{0}$ is mapped onto a final state 
$\ket{\Psi(T)}=U(T)\ket{\Psi(0)}$ lying also in $\mathscr{H}_{0}$. Such an 
iso-degenerate mapping is nothing but the adiabatic condition~\cite{Wilzeck}. 

Returning to the question of time evolution, we now seek an explicit 
representation of the final state $\ket{\Psi(T)}$. By applying the 
adiabatic condition, Eq.~(\ref{eq:8}) takes the form
\begin{equation}
\label{eq:9}
\mi\mathrm{D}_{t}\ket{\Psi(t)}=E_{0}(t)\Pi_{0}(t)\ket{\Psi(t)},
\end{equation}    
where $\Pi_{0}(t)=\sum_{a=1}^{n_{0}}\ket{\phi_{0}^{a}(t)}\bra{\tilde{\phi}_{0}^{a}(t)} $ 
for times $t\in[0,T]$ is the (pseudo-Hermitian) ground-state projector 
and $E_{0}$ denotes the lowest eigenvalue of $H$. As the state is initially 
prepared in $\mathscr{H}_{0}$ and will stay there while the evolution takes 
place, we can expand it in terms of the basis 
$\{\ket{\phi_{0}^{a}(t)}\}_{a=1}^{n_{0}}$, i.e.
\begin{equation}
\label{eq:10}
\ket{\Psi(t)}=\sum_{a=1}^{n_{0}}c_{a}(t)\ket{\phi_{0}^{a}(t)},
\end{equation}
with complex expansion coefficients $c_{a}(t)$. Inserting the expansion 
(\ref{eq:10}) into Eq.~(\ref{eq:9}) it is readily shown that
\begin{equation}
\label{eq:11}
\mi\sum_{a=1}^{n_{0}}(\dot{c}_{a}\ket{\phi_{0}^{a}}+c_{a}\ket{\dot{\phi}_{0}^{a}})
=\sum_{a=1}^{n_{0}}c_{a}(E_{0}\ket{\phi_{0}^{a}}+\mi K\ket{\phi_{0}^{a}}),
\end{equation}
where we used the definition of the covariant derivative $\mathrm{D}_{t}$. 

Contracting both sides of Eq.~(\ref{eq:11}) with $\bra{\tilde{\phi}_{0}^{b}}$ 
and noting that $\braket{\tilde{\phi}_{0}^{b}|\phi_{0}^{a}}=\delta_{ba}$, one 
obtains
\begin{equation*}
\label{eq:12}
\mi\dot{c}_{b}+\mi\sum_{a=1}^{n_{0}}c_{a}\braket{\tilde{\phi}_{0}^{b}|\dot{
\phi}_{0}^{a}}=E_{0}c_{b}+\mi\sum_{a=1}^{n_{0}}c_{a}\bra{\tilde{\phi}_{0}^{b}}K\ket{
\phi_{0}^{a}},
\end{equation*}
which can be rearranged as
\begin{equation}
\label{eq:13}
\dot{c}_{b}+\mi E_{0}c_{b}
+\sum_{a=1}^{n_{0}}c_{a}\bra{\tilde{\phi}_{0}^{b}}
\mathrm{D}_{t}\ket{\phi_{0}^{a}}=0.
\end{equation}
A formal solution to Eq.~(\ref{eq:13}) can be given in terms of a time-ordered 
integral~\cite{Bohm}. By introducing
$\left(A_{t}\right)^{ba}=\bra{\tilde{\phi}_{0}^{b}}\mi\mathrm{D}_{t}\ket
{\phi_{0}^{a}}$, a solution to Eq.~(\ref{eq:13}) is 
\begin{equation}
\label{eq:14}
c_{b}(T)=\sum_{a=1}^{n_{0}}\left[\boldsymbol{\hat{\mathrm{T}}}\mathrm{exp}
\int\limits_{0}^{T}\left[-\mi E_{0}(t)\mathds{1}+\mi A_{t}(t)\right]
\mathrm{d}t\right]^{ba}c_{a}(0).
\end{equation} 

An evolution in time is associated with a path 
$\gamma:[0,T]\rightarrow\mathscr{M}$ in a control manifold of the underlying 
quantum system. The $d$-dimensional manifold $\mathscr{M}$ is (locally) 
parametrized by a set of coordinates 
$\bm{\lambda}=\{\lambda^{\mu}\}_{\mu=1}^{d}$. These are the so-called control 
fields which drive the evolution of the Hamiltonian, i.e. 
$H(\bm{\lambda})=H_{\gamma(t)}$. In this framework, the time ordering for the 
integral over $A_{t}$ can be replaced by a path ordering 
$\boldsymbol{\hat{\mathrm{P}}}$ with respect to the parametrization by the 
coordinate chart $\{\lambda^{\mu}\}_{\mu=1}^{d}$.

Inserting the solution for the coefficients~(\ref{eq:14}) into the expansion 
(\ref{eq:10}), we find an explicit form for the quantum state after its 
evolution
\begin{gather}
\ket{\Psi(T)}=\sum_{a,b=1}^{n_{0}}\ket{\phi_{0}^{a}(0)}
\exp\left[-\mi\int\limits_{0}^{T}E_{0}(t)\mathrm{d}t\right]\nonumber\\
\times\left[\boldsymbol{\hat{\mathrm{P}}}\exp\left(\mi
\int\limits_{\bm{\lambda}(0)}^{\bm{\lambda}(T)}A\right)\right]^{ba}c_{a}(0),
\label{eq:15}
\end{gather}
where we introduced the gauge field (local connection one-form) 
$A=\sum_{\mu=1}^{d}A_{\mu}\mathrm{d}\lambda^{\mu}$. Its matrix-valued components 
$A_{\mu}$ 
are given by 
\begin{equation}
\label{eq:16}
\left(A_{\mu}\right)^{ba}=\mi\bra{\tilde{\phi}_{0}^{b}(\bm{\lambda})}
\left(\partial/\partial\lambda^{\mu}-K_{\mu}(\bm{\lambda})\right)\ket{\phi_{0}^{
a}(\bm{\lambda})},
\end{equation}
with 
$K_{\mu}(\bm{\lambda})=-\eta^{-1}(\bm{\lambda})\partial_{\mu}\eta(\bm
{\lambda})/2$ ($\partial_{\mu}=\partial/\partial\lambda^{\mu}$). Note that the 
components in Eq.~(\ref{eq:16}) contain a part that can 
be found in conventional QM and a metric-dependent term $K_{\mu}$. This has 
already been observed in Refs.~\cite{Wang,Gong2019} for the Abelian case. One recovers the Abelian result by setting $a=b$ and simplifying Eq.~(\ref{eq:16}) using
$\bra{\tilde{\phi}_{0}^{a}}K_{\mu}\ket{\phi_{0}^{a}}=[\bra{\tilde{\phi}_{0}^{a}}
\partial_{\mu}\ket{\phi_{0}^{a}}+\partial_{\mu}(\bra{\phi_{0}^{a}})\ket{\tilde{
\phi}_{0}^{a}}]/2$. In this notation 
$\left(A_{\mu}\right)^{aa}=-\mathfrak{I}\bra{\tilde{\phi}_{0}^{a}}\partial_{\mu}
\ket{\phi_{0}^{a}}$.

It can be straightforwardly shown that under a pseudo-unitary transformation
$\ket{\psi_{0}^{a}}=\sum_{c=1}^{n_{0}}U_{ca}\ket{\phi_{0}^{c}}$, the components 
of $A_{\mu}$ transform like a proper gauge field (cf. Appendix~\ref{sec:gauge}).
Furthermore the term $\mi A_{\mu}$ obeys a generalized anti-Hermiticity condition, that is
$[\mi(A_{\mu})^{ba}]^{\ast}=-\mi(A_{\mu})^{ab}_{\phi\leftrightarrow\tilde{\phi}}$,
where $\phi\leftrightarrow\tilde{\phi}$ means an interchanging of $\ket{\phi_{0}^{a}}$
and $\ket{\phi_{0}^{b}}$ by $\ket{\tilde{\phi}_{0}^{a}}$
and $\ket{\tilde{\phi}_{0}^{b}}$ respectively. The condition was derived by noting that 
$\bra{\tilde{\phi}_{0}^{a}}\partial_{\mu}\ket{\phi_{0}^{b}}=-\partial_{\mu}(\bra{\tilde{\phi}_{0}^{a}})\ket{\tilde{\phi}_{0}^{b}}$. 

The appearance of a gauge field in 
non-Hermitian QM was of course expected, as we started from a connection 
$\mathscr{H}=\mathscr{H}_{\mathrm{exc}}\oplus\mathscr{H}_{0}$. Note that
our gauge field differs from the one derived in \cite{Moste}, not only by 
a Lie-Algebra factor $\mi$ but also by the term $\bra{\tilde{\phi}_{0}^{a}}\partial_{\mu}\ket{\phi_{0}^{b}}$.

Turning back to Eq.~(\ref{eq:15}) and assuming that the state $\ket{\Psi}$ 
returns after a full period into its initial state up to a 
pseudo-unitary
rotation, $\ket{\Psi(0)}\rightarrow\ket{\Psi(T)}$, where the initial state is 
assumed to be one of the eigenstates $\ket{\phi_{0}^{a}(0)}$ rather than a 
superposition of them, we find
\begin{gather}
\ket{\Psi(T)}=\ket{\phi_{0}^{a}(0)}
\exp\left[-\mi\int\limits_{0}^{T} E_{0}(t)\mathrm{d}t\right] \nonumber\\
\times\sum_{b=1}^{n_{0}}\left[\mathcal{U}_{A}(\gamma)\right]^{ba}
c_{a}(0),
\label{eq:17}
\end{gather}  
where the cyclic time evolution corresponds to a loop $\gamma(0)=\gamma(T)$ in 
the parameter space $\mathscr{M}$. The mapping of the initial state 
$\ket{\phi_{0}^{a}(0)}$ described by Eq.~(\ref{eq:17}) is nothing but a unitary 
transformation with respect to the modified inner product 
$\braket{\cdot,\cdot}_{\eta}$. The exponential factor in Eq.~(\ref{eq:17}) 
is a dynamical phase factor, while the second term
\begin{equation}
\label{eq:18}
\mathcal{U}_{A}(\gamma)=\boldsymbol{\hat{\mathrm{P}}}\mathrm{exp}\left(\mathrm{
i}\oint\limits_{\gamma} A\right)
\end{equation} 
has purely geometric origin and is indeed a holonomy.

\section{Construction of pseudo-Hermitian systems}
\label{sec:theory}

For the purpose of illustration we shall consider a
benchmark Hamiltonian on which the previously devel-
oped theory can be studied.

\begin{figure}[h]
\centering
\includegraphics[width=6.5cm]{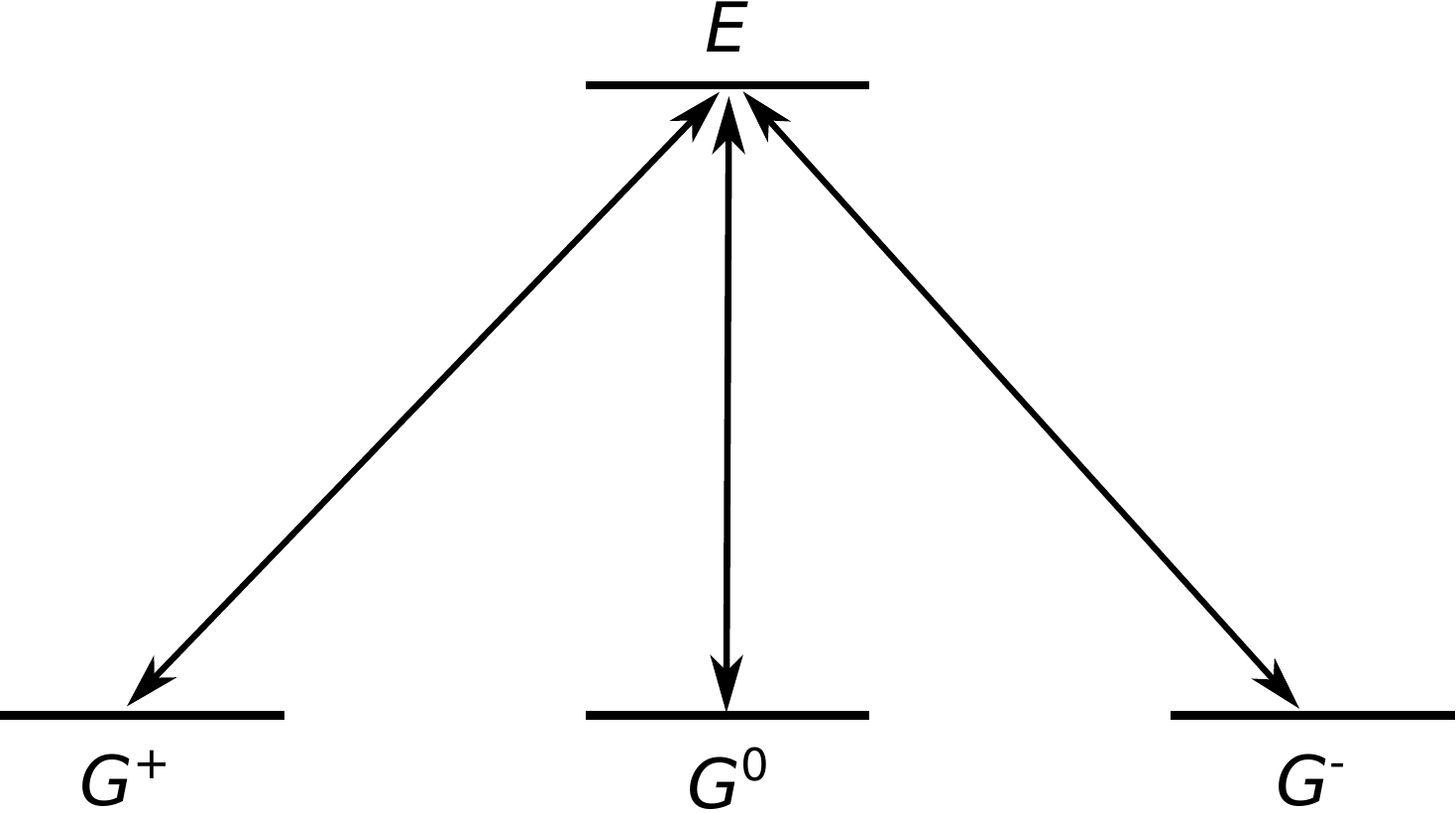}	
\caption{\label{fig:stirap} Representation of the level scheme of the 
pseudo-Hermitian Hamiltonian from Eq.~(\ref{eq:20}) in the time-varying Hilbert 
space $\mathscr{H}_{\eta(t)}$. In $\mathscr{H}_{\eta(t)}$ the Hamiltonian 
describes a tripod system.}
\end{figure}

There are mainly two approaches to construct artificial pseudo-Hermitian 
systems. The first route is to implement pseudo-Hermiticity via a \textit{top-down} approach in a
gain/loss system. For that one usually starts with an effective non-Hermitian Hamiltonian $H$ describing an open system phenomenologically.
The eigenvectors of this non-Hermitian Hamiltonian result directly in a biorthogonal basis as used in the previous sections.
This approach has the advantage that it is directly connected to a physical system.
For example, typical experimental realizations exist in the realm of optics, where the 
similarity of the paraxial Helmholtz equation with the Schr\"odinger equation 
allows one to design non-Hermitian characteristics with lossy waveguide systems 
\cite{Szameit1,Szameit2}. An approach using parity-time-symmetric lasing in an optical 
fibre network has been pursued in Ref.~\cite{Jahromi}, and in parity-time 
synthetic photonic lattices in Ref.~\cite{Regensburger}.
The second approach to non-Hermitian quantum theory is provided by biorthogonal 
quantum mechanics~\cite{Brody}. Given any biorthogonal basis, one can construct 
different pseudo-Hermitian systems from a \textit{bottom up} approach~\cite{Brody2}. 
Let us investigate the relation between these two approaches in more detail by 
considering a benchmark system. In the following, $H=L-\mi\Gamma$ is a 
complex Hamiltonian, with $L$ and $\Gamma$ being Hermitian operators 
given by
\begin{widetext}
\begin{equation*}
L=\frac{1}{2\Omega}\begin{pmatrix}
0&(\Omega-\Delta)\kappa_{0}^\ast&(\Delta+\Omega)\kappa_{-}
+(\Delta-\Omega)\kappa_{-}^\ast&(\Delta-\Omega)\kappa^\ast_{+}\\
(\Omega-\Delta)\kappa_{0}&0&(\Delta+\Omega)\kappa_{+}&0\\
(\Delta+\Omega)\kappa_{-}^\ast+(\Delta-\Omega)\kappa_{-}
&(\Delta+\Omega)\kappa_{+}^\ast&0&\frac{\alpha^2\kappa_{0}^\ast}{\Delta-\Omega}
\\
(\Delta-\Omega)\kappa_{+}&0&\frac{\alpha^2\kappa_{0}}{\Delta-\Omega}&0\\
\end{pmatrix},
\end{equation*}
\end{widetext}
\begin{equation*}
\Gamma=\frac{\alpha}{2\Omega}\begin{pmatrix}
|\kappa_{-}|^2&\kappa_{+}^\ast&0&\kappa_{0}^\ast\\
\kappa_{+}&0&\kappa_{0}&0\\
0&\kappa_{0}^\ast&-|\kappa_{-}|^{2}&-\kappa_{+}^\ast\\
\kappa_{0}&0&-\kappa_{+}&0\\
\end{pmatrix},\\
\end{equation*}  
where $\Omega(\alpha)=\sqrt{\Delta^2-\alpha^2}$ with $\Delta$ being a real 
constant and time-dependent parameters $\alpha(t)\in\mathds{R}$, 
$\kappa_{c}(t)\in\mathds{C}$. We assume that $0<\alpha^2<\Delta^2$ so that 
$\Omega$ stays real-valued. We can decompose $H$ as
\begin{equation}
\label{eq:20}
H=\sum_{c=0,\pm} \left( \kappa_{c}\ket{G^{c}(\alpha)}\bra{\tilde{E}(\alpha)} 
+\kappa_{c}^\ast\ket{E(\alpha)}\bra{\tilde{G}^{c}(\alpha)} \right),
\end{equation}
where 
\begin{equation*}
\ket{E}=\mathcal{N}_1
\begin{pmatrix}
\mi(\Omega-\Delta)\\
0\\
\alpha\\
0\\
\end{pmatrix},~\ket{G^{0}}=\mathcal{N}_1
\begin{pmatrix}
0\\
\mi(\Omega-\Delta)\\
0\\
\alpha\\
\end{pmatrix},
\end{equation*}
and 
\begin{equation*}
\ket{G^{-}}=\mathcal{N}_2
\begin{pmatrix}
-\mi(\Omega+\Delta)\\
0\\
\alpha\\
0\\
\end{pmatrix},~\ket{G^{+}}=\mathcal{N}_2
\begin{pmatrix}
0\\
-\mi(\Omega+\Delta)\\
0\\
\alpha\\
\end{pmatrix},
\end{equation*}
with normalization factors $\mathcal{N}_1=1/\sqrt{2\Omega(\Delta-\Omega)}$ 
and $\mathcal{N}_2=\mi/\sqrt{2\Omega(\Delta+\Omega)}$. Together with the 
associated states $\ket{\tilde{E}}=\ket{E}^\ast$ and $\ket{\tilde{G}^{c}}=\ket{G^{c}}^\ast$
for $c=0,\pm$, they form a biorthogonal basis. The Hamiltonian $H$ in Eq.~(\ref{eq:20}) 
possesses a two-fold degenerate dark subspace (zero eigenvalue eigenspace) and 
is therefore suitable for generating a pseudo-unitary, holonomic gate.
The Hamiltonian $H$ is the pseudo-Hermitian analogue of a typical 
light-matter coupling Hamiltonian that can be found in a variety of physical 
applications. 
For instance, in semiconductor quantum dots~\cite{QEC}, trapped ions~\cite{Duan}, or neutral atoms~\cite{Recati}. 
They all fall into the class of tripod systems.
By considering a controlled driving of the coupling parameters 
$\kappa_{c}=\kappa_{c}(t)$ a generalized STIRAP (Stimulated Raman adiabatic 
passage) process is induced \cite{Bergmann}. The system described by $H$ can 
therefore be seen as such a process taking place in a Hilbert space with a 
varying inner product structure $\braket{\cdot,\cdot}_{\eta(t)}$ (cf. 
Fig.~\ref{fig:stirap}). 

In the following we show that the Hamiltonian in Eq.~(\ref{eq:20}) can indeed be 
traced back to a Hermitian system. In general a Hermitian Hamiltonian $h$ can be expanded 
in an orthonormal basis $\{\ket{g^{c}},\ket{e}\}$, that is,
$\braket{g^{c}|g^{d}}=\delta_{cd}$, $\braket{g^{c}|e}=0$ and $\braket{e|e}=1$. 
This basis is related to the non-orthogonal states $\{\ket{E},\ket{G^{c}}\}$ by 
a generally non-unitary matrix $u$, i.e. $\ket{g^{c}}=u\ket{G^{c}}$ and 
$\ket{e}=u\ket{E}$. Similarly, for the associated states we have 
$\ket{g^{c}}=v\ket{\tilde{G}^{c}}$ and $\ket{e}=v\ket{\tilde{E}}$, where $v$ is 
some non-singular matrix. By construction, we have
\begin{gather}
\delta_{cd}=\braket{g^{c}|g^{d}}=\bra{\tilde{G}^{c}}v^\dagger u\ket{G^{d}},\nonumber\\
0=\braket{g^{c}|e}=\bra{\tilde{G}^{c}}v^\dagger u\ket{E},\nonumber\\
1=\braket{e|e}=\bra{\tilde{E}}v^\dagger u\ket{E},\nonumber
\end{gather}
only if $v^\dagger u=\mathds{1}$. 

We shall assume that $v\neq u$ to ensure that the problem is non-trivial. A 
relation of $u$ and $v$ to the metric operator $\eta$ is readily obtained. For 
example, starting from the state $\ket{E}$ we find that
\begin{equation*}
1=\braket{\tilde{E}|E}=\bra{E}\eta\ket{E}=\bra{e}v\eta u^{-1}\ket{e},
\end{equation*}
hence, $\eta=v^{-1}u=u^\dagger u$. Finally, we observe that the Hermitian 
counterpart $h$ to $H$ is given by $h=uHv^\dagger$. In the particular case of 
the Hamiltonian (\ref{eq:20}) we find
\begin{equation*}
h=\sum_{c=0,\pm} \left(\kappa_{c}\ket{g^{c}}\bra{e} 
+\kappa_{c}^\ast\ket{e}\bra{g^{c}} \right).
\end{equation*}

There is an additional remark to be made about the Hermitian system $h$. As 
pointed out in Ref.~\cite{Moste2}, there are a number of different 
representations of Hermitian counterparts of pseudo-Hermitian Hamiltonians. 
Especially, the Hamiltonian $\tilde{h}=\eta H$ is Hermitian as long 
as the spectrum of $H$ is real-valued, which can be seen from Eq.~(\ref{eq:1}). 
However, $\tilde{h}$ is then represented in a non-orthogonal basis.
A transformation to $h$, which is expanded in an orthonormal 
basis, is given by $\tilde{h}=(u^\dagger u)(v^\dagger hu)=u^\dagger hu$.
Which Hermitian analogue is suited to a specific system depends crucially on 
the basis in which one measures physical observables. 

\section{Evolution in dark subspaces}
\label{sec:dark}

We now turn to the Hamiltonian $H$ from Eq.~(\ref{eq:20}) to investigate its 
dynamics under an adiabatic evolution. At this point, one should recall that 
the metric operator of a pseudo-Hermitian system is in general not unique. It 
is well possible that a whole class of pseudo-Hermitian Hamiltonians is 
Hermitian under a certain metric operator. There might be even a 
time-independent metric under which $H$ is Hermitian. In order to resolve this 
ambiguity, we demand that the metric under which the observable $H$ is 
Hermitian, is the \textit{proper} metric $\eta$ given by the dyadic products of 
the left-handed eigenstates of $H$ \cite{Bender2}.  

We now investigate the dynamics induced by the Hamiltonian in Eq.~(\ref{eq:20}) 
with the aim to compute a holonomy. To do so, we have to consider a cyclic 
time evolution or, equivalently, a closed loop $\gamma$ in the parameter space 
$\mathscr{M}$. The evolution is assumed to be driven adiabatically by the 
time-dependence of the parameters $\kappa_{c}=\kappa_{c}(t)$. The holonomy will 
be generated in the degenerate dark subspace 
$\mathscr{H}_{\mathcal{D}}=\mathrm{span}\{\ket{\mathcal{D}^{l}}\}$. This is 
suitable for our computational purposes, as it neglects the uncontrollable 
dynamical phase ($E_{\mathcal{D}}(t)=0$ for all $t$). Throughout the dynamical 
process the parameter $\alpha$ will be assumed to be constant. As we will see 
this will reduce the computational effort by a lot. 

We seek a complete set of single qubit gates, thus ensuring that any 
pseudo-unitary gate with respect to the metric $\eta$ can be implemented 
over the dark subspace. It is sufficient to design a pair of non-commuting 
single-qubit gates $U_{1}$ and $U_{2}$. For the gate $U_{1}$ we choose the 
parametrization $\kappa_{-}=0$, 
$\kappa_{+}=-\kappa\sin(\vartheta/2)\mathrm{e}^{\mi\varphi}$ and 
$\kappa_{0}=\kappa\cos(\vartheta/2)$. In this case, the dark states are
\begin{gather}
\ket{\mathcal{D}^{1}}=\ket{G^{-}},\nonumber\\
\ket{\mathcal{D}^{2}}=
\cos(\vartheta/2)\ket{G^{+}}+\sin(\vartheta/2)\mathrm{e}^{\mi\varphi}\ket{G^{0}}
.
\label{eq:21}
\end{gather}
The remaining bright states (with eigenvalues $\pm\kappa$) read
\begin{gather}
\ket{\mathcal{B}^{+}}=\frac{1}{\sqrt{2}}\left(\sin(\vartheta/2)\ket{G^{+}}
-\mathrm{e}^{\mi\varphi}\cos(\vartheta/2)\ket{G^{0}}+\mathrm{e}^{\mi\varphi}\ket
{E}\right),\nonumber\\	
\ket{\mathcal{B}^{-}}=\frac{1}{\sqrt{2}}\left(\sin(\vartheta/2)\ket{G^{+}}
-\mathrm{e}^{\mi\varphi}\cos(\vartheta/2)\ket{G^{0}}-\mathrm{e}^{\mi\varphi}\ket
{E}\right).
\label{eq:22}
\end{gather}
Using the left-handed eigenstates associated to Eqs.~(\ref{eq:21}) and (\ref{eq:22}) we compute the full metric 
operator

\begin{gather}
\eta=\sum_{a=1,2}\ket{\tilde{\mathcal{D}}^{a}}\bra{\tilde{\mathcal{D}}^{a}}
+\ket{\tilde{\mathcal{B}}^{+}}\bra{\tilde{\mathcal{B}}^{+}}+\ket{\tilde{\mathcal
{B}}^{-}}\bra{\tilde{\mathcal{B}}^{-}},\nonumber\\
=\ket{\tilde{E}}\bra{\tilde{E}}+\sum_{c=0,\pm}\ket{\tilde{G}^{c}}\bra{\tilde{G}
^{c}}.
\label{eq:23}
\end{gather}
We recognize that as long as $\alpha$ stays constant, the metric operator 
$\eta$ does not depend on the parametrization of $\mathscr{M}$. In terms of the 
geometry of the underlying Hilbert space, a change of the parameter $\alpha$ 
leads to a contribution of the connection $K_{\mu}$. Hence, for 
$\alpha=\mathrm{const.}$ we have $K_{\mu}=0$. Thus, the gauge field 
(\ref{eq:16}) reduces to
\begin{equation*}
\left(A_{\mu}\right)^{ab}=\mi\bra{\tilde{\mathcal{D}}^a}\partial_{\mu}
\ket{\mathcal{D}^b}.
\end{equation*}
Evaluating the gauge field with respect to the coordinates 
$\lambda^{\mu}\in\{\vartheta,\varphi\}$ of $\mathscr{M}$, we get 
$\left(A_{\varphi}\right)^{22}=-\sin^2(\vartheta/2)$ as the only non-vanishing 
component of $A$. With this, we can compute the associated holonomy, and 
express the gate $U_{1}(\gamma)$ in terms of the Pauli matrices 
$\{\sigma^{x},\sigma^{y},\sigma^{z}\}$ with respect to the basis of dark states 
$\{\ket{\mathcal{D}^{1}},\ket{\mathcal{D}^{2}}\}$, viz.
\begin{equation}
\label{eq:27}
U_{1}(\gamma)=\mathrm{e}^{\mi\beta_{1}(\gamma)\ket{1}\bra{\tilde{1}}},
\end{equation}
where 
$\beta_{1}(\gamma)=-\oint_{\gamma}\sin^2(\vartheta/2)\mathrm{d}\vartheta\mathrm
{d}\varphi$.
Note that our computational basis is 
$\ket{0}=\ket{\mathcal{D}^{1}(\bm{0})}=\ket{G^{-}}$ and 
$\ket{1}=\ket{\mathcal{D}^{2}(\bm{0})}=\ket{G^{+}}$.
In Eq.~(\ref{eq:27}), path-ordering can be neglected, as the chosen 
parametrization effectively generates an Abelian geometric phase, i.e. the 
matrix-valued components $A_{\vartheta}$ and $A_{\varphi}$ commute.

For the second gate $U_{2}$, we choose the parametrization 
$\kappa_{0}=\kappa\cos(\vartheta)$, 
$\kappa_{-}=\kappa\sin(\vartheta)\cos(\varphi)$ and 
$\kappa_{+}=\kappa\sin(\vartheta)\sin(\varphi)$, and repeat the previous 
calculation in a similar fashion, starting with the new dark states
\begin{gather}
 \ket{\mathcal{D}^{1}}=\cos(\vartheta)\left[\cos(\varphi)\ket{G^{-}}
+\sin(\varphi)\ket{G^{+}}\right]-\sin(\vartheta)\ket{G^{0}},\nonumber\\
\ket{\mathcal{D}^{2}}=\cos(\varphi)\ket{G^{+}}-\sin(\varphi)\ket{G^{-}}.\nonumber
\end{gather}
Together with the associated bright states we obtain in this case the same 
metric operator as in Eq.~(\ref{eq:23}). Hence, $K_{\mu}=0$ as long as 
$\alpha\neq\alpha(t)$. We find the components of the gauge field in 
$\mathscr{H}_{\mathcal{D}}$ to be
\begin{equation}
\label{eq:29}
A_{\vartheta}=0, \quad A_{\varphi}=\cos(\vartheta)\sigma^{y},
\end{equation}
so that path-ordering can be neglected again. 

The associated holonomy $U_{2}$ to $A$ is thus given by inserting 
Eq.~(\ref{eq:29}) into Eq.~(\ref{eq:18}). Explicitly we have
\begin{equation*}
U_{2}(\gamma^\prime)=\mathrm{e}^{\mi\beta_{2}(\gamma^\prime)\sigma^{y}},
\end{equation*}
where 
$\beta_{2}(\gamma^\prime)=\oint_{\gamma^\prime}\cos(\vartheta)\mathrm{d}
\vartheta\mathrm{d}\varphi$ for a path $\gamma^\prime$ in $\mathscr{M}$.
From here, one is able to compute the commutator of $U_{1}$ and $U_{2}$, that is
\begin{equation}
\label{eq:31}
[U_{1},U_{2}]=\sin(\beta_{2})\left(1-\mathrm{e}^{\mi\beta_{1}}\right) \sigma^{x}.
\end{equation}
In general, Eq.~(\ref{eq:31}) does not vanish for generic loops $\gamma$ and 
$\gamma^\prime$. Hence, we have found a universal set of pseudo-unitary 
single-qubit gates on which HQC could be based. This is the key result of this 
work.
The presented procedure shows how a lossy system, which generates the
Hamiltonian for a generic holonomic computation, can
be described effectively in the pseudo-Hermitian picture.
The range of new applications that could stem from this extension of the theory 
needs further investigations and is out of the scope of this work.

\section{Discussion and concluding remarks}
\label{sec:conclusions}

In this article we have shown how the holonomic approach to quantum computation can be extended to
pseudo-Hermitian systems. We derived a non-Abelian
geometric phase generating a pseudounitary holonomy
over the degenerate eigenspace. The gauge field associated with the non-Abelian phase contains an additional term due to
the modified inner product structure induced by a pseudo-Hermitian quantum system, which is absent
in conventional quantum mechanics.

This general framework was applied to a benchmark
Hamiltonian that can be implemented in terms of a
gain/loss system. By choice of a suitable biorthogonal
basis the system has the form of a tripod Hamiltonian. 
An explicit calculation showed that the considered system allows for the
implementation of a arbitrary pseudounitary transformations over the two-dimensional dark subspace.

Furthermore, we investigated the underlying geometry
of this Hamiltonian. In particular, we have shown
that the inner product structure could be held constant
throughout an adiabatic evolution. This can be done by
choosing a suitable loop in the parameter space such that
the additional term, appearing in the geometric phase,
vanishes. Therefore this loop only changed the coupling
between certain tripod levels but does not involve the
biorthogonal basis, i.e. the inner product structure, in
which the Hamiltonian is represented.
Generalized to arbitrary pseudo-Hermitian systems, this enables clear
analysis of pseudounitary holonomies and their dependence
on the changing inner product structure.

\acknowledgments
Financial support by the Deutsche Forschungsgemeinschaft (DFG SCHE 612/6-1) 
is gratefully acknowledged.

\appendix

\section{Transformation law for the gauge field}
\label{sec:gauge}

Here we show that $A$ indeed transforms like a proper gauge 
field \cite{Nakahara} under a change of basis 
$\ket{\psi^{a}}=\sum_{i=1}^{n}U_{ia}\ket{\phi^{i}}$, where 
$U_{ia}\in\mathds{C}$. The transformation is mediated by a pseudo-unitary matrix
\begin{equation*}
\mathcal{U}(\bm{\lambda})=\sum_{i,j=1}^{n}U_{ij}(\bm{\lambda})\ket{\phi^{i}(\bm{
\lambda})}\bra{\tilde{\phi}^{j}(\bm{\lambda})}\in\mathrm{U}_{\eta}(n_{0}).
\end{equation*}
Here, $\mathrm{U}_{\eta}(n)$ is the group of $n$-dimensional 
$\eta$-pseudo-unitary matrices \cite{Moste3}. We find the usual transformation 
law
\begin{gather}
\left(A_{\mu}^\prime\right)^{ab}
=\mi \bra{\tilde{\psi}^{a}}
\left(\partial_{\mu}-K_{\mu}\right)\ket{\psi^{b}} 
\nonumber \\
=\mi\sum_{i,j=1}^{n}\bra{\phi^{i}}U_{ia}^\ast\eta
\left(\partial_{\mu}-K_{\mu}\right)U_{jb}\ket{\phi^{j}} 
\nonumber \\
=\mi\sum_{i,j=1}^{n}\left(\bra{\phi^{i}}U_{ia}^\ast\eta\left(\partial_{\mu}U_{bj}
\right)\ket{\phi^{j}}\right. \nonumber \\
\left.+\bra{\phi^{i}}U_{ia}^\ast\eta U_{bj}\partial_{\mu}\ket{\phi^{j}}
-\bra{\phi^{i}}U_{ia}^\ast\eta K_{\mu}U_{bj}\ket{\phi^{j}}\right) 
\nonumber \\
=\sum_{i,j=1}^{n}U_{ia}^\ast\mi\partial_{\mu}U_{bj}\delta_{ij}
+U_{ia}^\ast U_{bj}\left(A_{\mu}\right)^{ij} \nonumber \\
=\sum_{i=1}^{n}U_{ia}^\ast\mi\partial_{\mu}U_{bi}
+\sum_{i,j=1}^{n}U_{ia}^\ast U_{bj}\left(A_{\mu}\right)^{ij},\nonumber
\end{gather}
or, in matrix notation,
\begin{equation*}
A_{\mu}\mapsto \mathcal{U}^{-1} 
 A_{\mu}\mathcal{U}+\mathcal{U}^{-1}\mi\partial_{\mu}\mathcal{U}.
\end{equation*}

\section{Natural geometric picture of pseudo-Hermitian Hamiltonians}
\label{sec:Grassmann}

So far, our treatment of pseudo-Hermitian Hamiltonians did not involve the 
language of fiber bundles. In conventional QM it is well known that the 
projector formalism used in HQC involves more advanced concepts such as 
Grassmann and Stiefel manifolds \cite{Fujii}. To the best of our knowledge, 
these notions have not been established for pseudo-Hermitian systems yet.

Let us consider a pseudo-Hermitian Hamiltonian $H\in\mathrm{End}(\mathscr{H})$, with
$R+1$ different eigenvalues, defined over the $N$-dimensional Hilbert space $\mathscr{H}$. 
Suppose $H$ has a real spectrum so that its spectral decomposition reads
\begin{equation*}
 H=\sum_{l=0}^{R}E_{l}\Pi_{l},
\end{equation*}
where $\{E_{l}\}_{l=0}^{R}$ are the eigenvalues corresponding to the 
pseudo-Hermitian projector 
$\Pi_{l}=\sum_{k=1}^{n_{l}}\ket{\phi_{l}^{k}}\bra{\tilde{\phi}_{l}^{k}}$ with 
$n_{l}$ being the degeneracy of the $l$-th level. The states 
$\{\ket{\phi_{l}^{k}}\}_{k=1}^{n_{l}}$ of the $l$-th eigenspace of $H$ form a 
biorthogonal frame
\begin{equation}
 \label{eq:41} 
V_{l}=\sum_{k=1}^{n_{l}}\ket{\phi_{l}^{k}}\bra{\tilde{k}}\cong\left(\ket{\phi_{
l}^{1}},\dots,\ket{\phi_{l}^{n_{l}}}\right)_{\ket{\tilde{k}}},
\end{equation}
where $\{\ket{\tilde{k}}\}_{k=1}^{n_{l}}\subset\mathds{C}^{n_{l}}$ constitutes a 
complete, biorthogonal basis with $\{\ket{k}\}_{k=1}^{n_{l}}$, where 
$\ket{\tilde{k}}=\eta_{\mathrm{a}}\ket{k}$. 
Note that this basis is of no physical relevance and acts merely as a tool to 
represent the frame $V_{l}$. One can indeed choose 
$\eta_{\mathrm{a}}=\mathds{1}_{n_{l}}$ so that $\{\ket{k}\}_{k=1}^{n_{l}}$ forms an 
orthonormal basis.

The notion of biorthogonal frames gives rise to a more subtle issue. Usually, 
in the study of pseudo-Hermitian and pseudo-unitary operators, one is 
confronted with square matrices. Because $V_{l}$ is not an 
observable, we have to modify the pseudo-Hermiticity condition~(\ref{eq:1})
for non-square matrices.
In the case of a biorthogonal frame, we can
define the pseudo-adjoint matrix of $V_{l}$ as
\begin{equation*}
V_{l}^{\ddagger}=\eta_{\mathrm{a}}^{-1}V_{l}^\dagger\eta,
\end{equation*}
where $\eta\in\mathds{C}^{N\times N}$ is the metric operator formed from the 
left-handed eigenstates $\ket{\tilde{\phi}_{l}^{k}}$, that is
\begin{equation*}
 \eta=\sum_{l=0}^{R}\sum_{k=1}^{n_{l}}\ket{\tilde{\phi}_{l}^{k}}\bra{\tilde{\phi}_{l}^{k}}.
\end{equation*}
Note that $\eta$ serves as a metric 
for the projector $\Pi_{l}$, i.e. 
\begin{equation*}
 \Pi_{l}^\dagger\eta=\Pi_{l}^\dagger\eta_{l}=\eta_{l}\Pi=\eta\Pi,
\end{equation*}
where $\eta_{l}=\sum_{k=1}^{n_{l}}\ket{\tilde{\phi}_{l}^{k}}\bra{\tilde{\phi}_{l}^{k}}$.

Representing the biorthogonal frame~(\ref{eq:41}) by a complex $(N\times 
n_{l})$-matrix we find its pseudo-adjoint to be
\begin{equation*}
V^{\ddagger}_{l}=\sum_{k=1}^{n_{l}}\ket{k}\bra{\tilde{\phi}_{l}^{k}}\cong
\begin{pmatrix}
\bra{\tilde{\phi}_{l}^{1}}\\ \vdots\\ 
\bra{\tilde{\phi}_{l}^{n_{l}}}\\ 
\end{pmatrix}\in\mathds{C}^{n_{l}\times N}.
\end{equation*}
By construction, we have $V_{l}^{\ddagger}V_{l}=\mathds{1}_{n_{l}}$, which verifies 
that the set $\{\ket{\phi_{l}^{k}}\}_{k=1}^{n_{l}}$ constitutes a biorthogonal 
basis for the ground-state eigenspace. The set of all biorthogonal frames is 
called the Stiefel manifold defined by
\begin{equation*}
 S_{N,n_{l},\eta}=\{V_{l}\in\mathds{C}^{N\times n_{l}}\,|\, 
V_{l}^{\ddagger}V_{l}=\mathds{1}_{n_{l}}\}.
\end{equation*}
It is noteworthy that the projector $\Pi_{l}$ can be expressed in terms of a 
biorthogonal frame in $S_{N,n_{l}}$ (we have dropped $\eta$ for ease of 
notation), i.e. $\Pi_{l}=V_{l}V_{l}^{\ddagger}$. It is easily checked that the 
so defined projector belongs to the Grassmann manifold
\begin{gather}
 G_{N,n_{l}}=\{\Pi_{l}\in\mathds{C}^{N\times N}\,|\,\Pi_{l}^{2}=\Pi_{l}, 
\nonumber\\
\Pi_{l}^{\ddagger}=\Pi_{l},\,\mathrm{tr}\left(\Pi_{l}\right)=n_{l}\}.\nonumber
\end{gather}
Because the projector is a square matrix, its pseudo-adjoint is defined in 
the usual sense \cite{Moste2},
$\Pi_{l}^{\ddagger}=\eta^{-1}\Pi_{l}^{\dagger}\eta$.

We are now in a position to illuminate the gauge freedom within the projector 
$\Pi_{l}$. More precisely, we can define a projection $\pi$ from the Stiefel 
manifold to the Grassmann manifold by $V_{l}\mapsto V_{l}V_{l}^{\ddagger}$. It 
is not hard to show that the image of this map stays invariant under a group 
action by a pseudo-unitary matrix 
$\mathcal{U}\in\mathrm{U}_{\eta_{\mathrm{a}}}(n_{l})$,
\begin{equation*} 
\pi(V_{l}\mathcal{U})=(V_{l}\mathcal{U})(V_{l}\mathcal{U})^{\ddagger}=V_{l}
\mathcal{U}\mathcal{U}^{\ddagger}V_{l}^{\ddagger}=V_{l}V_{l}^{\ddagger},
\end{equation*}
where we applied the useful relation 
\begin{equation*}
(V_{l}\mathcal{U})^{\ddagger}=\eta_{\mathrm{a}}^{-1}(V_{l}\mathcal{U})^{\dagger}
\eta=\eta_{\mathrm{a}}^{-1}\mathcal{U}^{\dagger}\left(\eta_{\mathrm{a}}\eta_
{\mathrm{a}}^{-1}\right)V_{l}^{\dagger}\eta=\mathcal{U}^{\ddagger}V_{l}^{
\ddagger}.
\end{equation*}

In conclusion we have constructed a 
$\mathrm{U}_{\eta_{\mathrm{a}}}(n_{l})$-principal bundle, that is,
\begin{equation}
 \label{eq:47}
 S_{N,n_{l}}\overset{\pi}{\longrightarrow}G_{N,n_{l}}.
\end{equation}
The bundle structure, Eq.~(\ref{eq:47}), is a direct generalization of the one 
found in conventional QM (for a review, see e.g. \cite{Fujii}). The standard 
theory is recovered for $\eta=\mathds{1}_{N}$. It is therefore not surprising 
that the Stiefel manifold can be written, in analogy to their counterparts in 
conventional QM, as a coset space, i.e.
\begin{equation*}
 G_{N,n_{l}}\cong S_{N,n_{l}}/\mathrm{U}_{\eta_{\mathrm{a}}}(n_{l}).
\end{equation*}
Note how, for $n_{l}=1$ (i.e. a non-degenerate situation), the Grassmann 
manifold reduces to the projective Hilbert space containing the pseudo-Hermitian 
density operators for a pure state, 
i.e. $\ket{\phi_{l}}\bra{\tilde{\phi_{l}}}\in G_{N,1}\cong\mathds{C}P^{N-1}$. 
The structure group of this principal bundle is 
$\mathrm{U}_{\eta_{\mathrm{a}}}(1)$ which is identical to the conventional 
unitary group $\mathrm{U}(1)$.

We conclude this section by recalling that it is rather demanding for a 
parameter space $\mathscr{M}$ to be mapped one-to-one (bijectively) onto 
$G_{N,n_{l}}$. In other words, a realistic quantum system, given by a family 
$\{H(\lambda)\}_{\lambda\in\mathscr{M}}$ of iso-spectral pseudo-Hermitian 
Hamiltonians, may have a smaller control manifold than the whole Grassmann 
manifold. Nevertheless, there is a map $\Phi$ from $\mathscr{M}$ onto 
$G_{N,n_{l}}$ defined by $\Phi(\lambda)=\Pi_{l}$. A natural way to 
study the geometry of such systems is given in terms of the pullback bundle of 
the Stiefel manifold
\begin{equation}
\label{eq:49}
\Phi^\ast S_{N,n_{l}}=\left\{(\lambda,V_{l})\in\mathscr{M}\times 
S_{N,n_{l}}\,\bigr|\,\,\pi_{\lambda}(V_{l}
)=\Phi(\lambda)\right\}.
\end{equation}
In order to construct the rest of the bundle structure of Eq.~(\ref{eq:49}), we 
can establish the fiber $\mathscr{F}_{\lambda}$ of $\Phi^{*}S_{N,n_{l}}$ over the point 
$\lambda$ in $\mathscr{M}$. This fiber is just a copy of the fiber 
$\mathscr{F}_{\Phi(\lambda)}$ defined over the projector (point in 
$G_{N,n_{l}}$) $\Pi_{l}$. The latter is formally defined as the 
preimage of the projection $\pi(V_{l})=\Pi_{l}$, that is,
$\mathscr{F}_{\Phi(\lambda)}\cong \pi^{-1}(\Pi_{l})$. Then 
\begin{equation*} 
\Phi^\ast S_{N,n_{l}}\overset{\pi_{\Phi}}{\longrightarrow}
\mathscr{M},
\end{equation*}
where $\pi_{\Phi}:(\lambda,V_{l})\mapsto\lambda\in\mathscr{M}$, constitutes a 
$U_{\eta_{\mathrm{a}}(\lambda)}(n_{l})$-principal fiber bundle. By construction,
the sections of this bundle are just $\lambda\mapsto(\lambda,V_{l})$.


\begin{thebibliography}{99}

\bibitem{Bender}
C. M. Bender and S. Boettcher, Phys. Rev. Lett. \textbf{80}, 5243 (1998).
\bibitem{Moste2} 
A. Mostafazadeh, Int. J. Geom. Meth. Mod. Phys. \textbf{07}, 1191 (2010).
\bibitem{Brody} 
D. C. Brody, J. Phys. A: Math. Theor. \textbf{47}, 035305 (2013).
\bibitem{HQC}
P. Zanardi and M. Rasetti, Phys. Lett. A \textbf{264}, 94 (1999).
\bibitem{Pachos2}
J. Pachos and P. Zanardi, Int. J. Mod. Phys. B \textbf{15}, 1257 (2001).
\bibitem{Berry}
M. V. Berry, Proc. Roy. Soc. A. Math. Phys. Sci. \textbf{392}, 45 (1984).
\bibitem{Wilzeck}
F. Wilczek and A. Zee, Phys. Rev. Lett. \textbf{52}, 2111 (1984). 
\bibitem{Dalibard}
J. Dalibard, F. Gerbier, G. Juzeli$\overline{\mathrm{u}}$nas, and P. {\"O}hberg, Rev. Mod. Phys. \textbf{83}, 1523 (2011).
\bibitem{Berger}
A. A. Abdumalikov Jr., J. M. Fink, K. Juliusson, M. Pechal,
S. Berger, A. Wallraff, and S. Filipp, Nature \textbf{496}, 482 (2013).
\bibitem{Teuber}
M. Kremer, L. Teuber, A. Szameit, and S. Scheel, arXiv:1902.02559v1 [physics.optics] (2019).
\bibitem{MaSO}
L. B. Ma, S. L. Li, V. M. Fomin, M. Hentschel, J. B. G{\"o}tte,
Y. Yin, M. R. Jorgensen, and O. G. Schmidt, Nat. Commun. \textbf{7}, 10983 (2016).
\bibitem{QEC}
D. A. Lidar and T. A. Brun, \textit{Quantum Error Correction} (Cambridge 
University Press, Cambridge, 2013).
\bibitem{Bohm}
A. Bohm, A. Mostafazadeh, H. Koizumi, Q. Niu, J. Zwanziger, \textit{The Geometric Phase in Quantum Systems} (Springer, Berlin, 2013).
\bibitem{Moste}
A. Mostafazadeh, Phys. Rev. D \textbf{98}, 046022 (2018).
\bibitem{Pachos}
J. Pachos and S. Chountasis, Phys. Rev. A \textbf{62}, 052318 (2000).
\bibitem{Romero}
Y. Wang, J. Zhang, C. Wu, J. Q. You, and G. Romero, Phys. Rev. A \textbf{94}, 012328 (2016).
\bibitem{Oreshkov}
O. Oreshkov, T. A. Brun, and D. A. Lidar, Phys. Rev. Lett. \textbf{102}, 070502 (2009).
\bibitem{Bhandari}
J. Samuel and R. Bhandari, Phys. Rev. Lett. \textbf{60}, 2339 (1988).
\bibitem{Wang0}
J. Gong and Q. Wang, Phys. Rev. A \textbf{82}, 012103 (2010).
\bibitem{Wang}
D.-J. Zhang, Q.-h. Wang, and J. Gong, arXiv:1811.04640v1 [quant-ph] (2018).
\bibitem{Gong2019}
D.-J. Zhang, Q.-h. Wang, and J. Gong, Phys. Rev. A \textbf{99}, 042104 (2019).
\bibitem{Brody2}
D. C. Brody, J. Phys. A: Math. Theor. \textbf{49}, 10LT03 (2016).
\bibitem{Wang2}
J. Gong and Q.-h. Wang, J. Phys. A: Math. Theor. \textbf{46}, 485302 (2013).  
\bibitem{Heiss}
W. D. Heiss, J. Phys. Math. Theor. \textbf{45}, 444016 (2012). 
\bibitem{Bender2}
C. M. Bender, D. C. Brody, and H. F. Jones, Amer. J. Phys. \textbf{71}, 
1095 (2003).
\bibitem{Szameit1} 
A. Szameit and S. Nolte, J. Phys. B: At. Mol. Opt. Phys. \textbf{43}, 163001 (2010).
\bibitem{Szameit2}
A. Szameit, M. C. Rechtsman, O. Bahat-Treidel, and M. Segev, Phys. Rev. A \textbf{84}, 021806(R) (2011).
\bibitem{Jahromi}
A. K. Jahromi, A. U. Hassan, D. N. Christodoulides, and A. F. Abouraddy, Nat. Commun. \textbf{8}, 1359 (2017).
\bibitem{Regensburger}
A. Regensburger, C. Bersch, M.-A. Miri, G. Onishchukov, D. N. Christodoulides 
and U. Peschel, Nature \textbf{488}, 167 (2012).
\bibitem{Duan}
L. M. Duan, J. Cirac, and P. Zoller, Science \textbf{292}, 1695 (2001).
\bibitem{Recati}
A. Recati, T. Calarco, and P. Zanardi, J. I. Cirac, and P. Zoller, Phys. Rev. A \textbf{66}, 032309 (2002).
\bibitem{Bergmann}
R. G. Unanyan, B. W. Shore, and K. Bergmann, Phys. Rev. A \textbf{59}, 2910 (1998).
\bibitem{Nakahara}
M. Nakahara, \textit{Geometry, Topology and Physics}, 2$^{\mathrm{nd}}$ edition (Taylor \& Francis, New York, 2013).
\bibitem{Moste3}
A. Mostafazadeh, J. Math. Phys. \textbf{45}, 932 (2004).
\bibitem{Fujii}
K. Fujii, Rep. Math. Phys. \textbf{48}, 75 (2001).

\end{thebibliography}
\end{document}